\documentclass[epj]{svjour}

\newcommand{\bea}{\begin{eqnarray}}
\newcommand{\eea}{\end{eqnarray}}
\newcommand{\spaceint}[2]{\int_{#1} d^3 #2 \;}
\newcommand{\vect}[1]{\mathbf{#1}}

\newcommand{\di}{\displaystyle}

\usepackage{amsmath,amssymb}
\usepackage{latexsym}
\usepackage{epsfig,color}
\usepackage[normalem]{ulem}

\begin{document}

%\newlength{\mylen}
%\setlength{\mylen}{\textwidth}
%\addtolength{\mylen}{-1cm}

%\preprint{{\LARGE \bf DRAFT}}

\title{Effective interactions of colloids on nematic films}

\author{M. Oettel\inst{1}, A. Dom\'\i nguez\inst{2}, M. Tasinkevych\inst{3}
 \and S. Dietrich\inst{3}}

\institute{
 Johannes--Gutenberg--Universit\"at Mainz, Institut f{\"ur} Physik,
  WA 331, D--55099 Mainz, Germany 
\and
 F\'\i sica Te\'orica, Universidad de Sevilla, Apdo.1065, E-41080
  Sevilla, Spain
\and 
 Max-Planck-Institut f\"ur Metallforschung, Heisenbergstr. 3,
  D-70569 Stuttgart and
 Institut f\"ur Theoretische und Angewandte Physik, Universit\"at Stuttgart,
             Pfaffenwaldring 57, D-70569 Stuttgart, Germany
}

\date{Received: / Revised version:}

\abstract{
  The elastic and capillary interactions  between a pair of colloidal particles
  trapped on top of a nematic film are studied theoretically 
  for large separations $d$. The elastic interaction is repulsive and 
  of quadrupolar type, varying as $d^{-5}$.
  For macroscopically thick films, the capillary interaction is likewise
  repulsive and proportional to $d^{-5}$ as a consequence
  of mechanical isolation of the system comprised of the colloids and the interface.
  A finite film thickness introduces a nonvanishing force on the system
  (exerted by the substrate supporting the film)
  leading to logarithmically varying capillary attractions. 
  However, their strength turns out to be too small to 
  be of importance for the recently observed pattern formation of 
  colloidal droplets on
  nematic films.
}

 % end of PACS codes

\maketitle

\section{Introduction}

The interactions of colloidal particles trapped at fluid interfaces have been found to differ
significantly from the corresponding interactions in bulk solvents. This has been studied
mostly for electrically charged particles trapped at interfaces with water. On 
one hand, the presence of the interface gives rise to direct dipolar electrostatic
repulsions between the colloids (see Refs.~\cite{Pie80,Ave02,Par07} for some  experimental evidence),
on the other hand deformations of the interface may induce longer--ranged
capillary attractions (briefly reviewed in Refs.~\cite{Zen06,Bre7r,Oet08}) which 
is possibly the source of 
pattern formation observed
in various experiments \cite{Ghe97,Gar98a,Que01,Gom05,Che06}. 
(See, however, Ref.~\cite{Fer04} for an alternative
explanation due to interface impurities.)

\begin{figure}[t]
 \begin{center}
  \epsfig{file=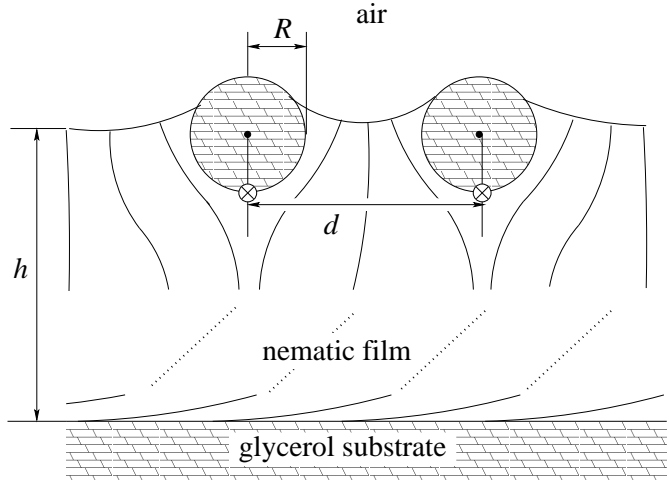, width=\columnwidth}
 \end{center}
 \caption{ \label{fig:nem_exp}
 Schematic setup of the experiment reported in Ref.~\cite{Lav04}. 
 Colloidal glycerol drops ($R = 1 \dots 7$ $\mu$m) are trapped at the surface of a   
 thick nematic film ($h \approx 60$ $\mu$m). The director field in the
 nematic film is sketched by the black lines and their dotted interpolations. 
 Nematic anchoring at the glycerol substrate at the bottom of the film
 as well as on the surface of the glycerol drops 
 is parallel, whereas it is perpendicular (homeotropic) 
 at the deformed nematic--air interface. Each
 drop is necessarily accompanied by a topological defect ($\otimes$). 
 }
\end{figure}

Recently \cite{Lav04,Lav07}, the experimental observation of ordered structures of glycerol 
droplets bound to a 
nematic--air interface has been reported and attributed to
an effective pair potential between the colloids which contains a repulsive,
elastic part due to director deformations in the supporting
nematic film and an attractive, capillary part which is long--ranged and mediated 
by logarithmically 
varying deformations of the nematic--isotropic interface caused by the droplets.   
The schematic setup of this experiment is depicted in
Fig.~\ref{fig:nem_exp}. According to Ref.~\cite{Lav04}, the colloidal particles 
experience an 
upward force caused by elastic forces due to director deformations in the supporting
nematic film. 
%and the resulting surface mediated colloid--colloid interaction can be evaluated 
%in the superposition approximation. 
This upward force on the colloids is supposed to give rise to
the aforementioned logarithmically varying  interface deformation.
Applying a superposition approximation for the
deformation field, one can show that the ensuing effective capillary interaction
potential between two colloids is likewise varying logarithmically. 
This is similar to the flotation interaction
of mm--sized particles at fluid interfaces for which the force on the colloids is caused
by gravity (see, e.g., Ref.~\cite{Kra00}) and also parallels the tentative 
explanation given for the experimentally observed attractions
between sub-$\mu$m charged colloids at a water-oil interface \cite{Nik02}
(for the controversy around this explanation see 
Refs.~\cite{Meg03,Nik03,For04,Kra04,Oet05,Oet05a,Kra05,Dan06}). 

However, it is now well established \cite{Meg03,Nik03,For04,Oet05,Oet05a,Kra05,Dom06}
%some of us have shown recently \cite{Oet05} 
that interface deformations and effective colloidal interactions varying logarithmically only arise in experimental systems which are not isolated
mechanically. For mechanically isolated systems it can be shown \cite{Oet05b,Wue05,Dom05,Dom07}
that both the interface deformation around  a single  colloid and the effective
interaction between two of them  are shorter--ranged and the latter cannot be 
calculated reliably within the
superposition approximation. 

In the following we will extend the arguments presented in 
Refs.~\cite{Oet05,Oet05b,Wue05,Dom05,Dom07}
to systems with colloids at nematic interfaces. We will show that mechanical isolation
of the system ``nematic film -- colloid -- air" can be violated through a subtle interplay
between the finite thickness of the film and the anchoring conditions at the colloids
and at the nematic interfaces with the substrate and with the  air, respectively. 
However, for 
experimental conditions as the ones described in 
Ref.~\cite{Lav04} a quantitative estimate of the strength of the ensuing 
logarithmic attraction between the colloids yields that these attractions
are unobservably small. Therefore it seems likely that this kind of asymptotic capillary forces
cannot be invoked as a relevant mechanism to account for the observations 
reported in Ref.~\cite{Lav04}.  

The manuscript is organized as follows:
In Sec.~\ref{sec:coarse} the coarse--grained model for the nematic
phase is introduced which will serve as the basis for all subsequent
calculations. In Subsec.~\ref{sec:thick} we study the case of an
infinitely thick nematic film. First, we compute the asymptotic form of
the director field and the ensuing elastic force between two
particles. Then we calculate the effective force arising from the
deformation of the fluid--nematic interface caused by the elastic
stresses. In Subsec.~\ref{sec:thin} we consider a nematic film of finite
thickness, which models more closely the experimental setup described
in Ref.~\cite{Lav04}; for such a system we extend the above calculations 
to the two opposite cases
of perpendicular and parallel anchoring of the director field at the
substrate surface.  In Sec.~\ref{sec:conclusion} we
discuss our results.

%that for a  
%mechanically isolated experimental system (which is reasonable to assume since
%gravity is unimportant for $\mu$m--sized particles) an acting force on the colloid
%is necessarily accompanied by a stress on the interface in such a way that 
%for small forces the
%resulting interface deformation is {\em not} logarithmic but short--ranged:
%if the stress field on the interface around one colloid $\Pi(r)$ decays like
%$r^{-n}$ the interface deformation $u(r)$ decays like $r^{-n+2}$.   
%Consequently, the superposition approximation predicts a short--ranged, 
%{\em repulsive} surface mediated potential $V_{\rm men}(d)$ for colloids at distance
%$d$ which varies as $d^{-n}$. Although we have investigated 
%only charged colloids in more detail, the arguments are quite general and can be applied
%to a nematic interface as well which we will demonstrate below. 

\section{Coarse--grained model}
\label{sec:coarse}

In view of the mesoscopic length scales involved we describe the bulk part of the 
nematic free energy associated with the director deformations
in terms of the Frank free energy expression within the one--coupling approximation
\cite{deG74}
\bea
  {\cal F}^{\rm b}_{\rm ne} &=&  \spaceint{V_{\rm ne}}{r} f^{\rm b} (\vect r) \; \nonumber \\
 \label{eq:f_lc}
           &=& \frac{K}{2}\spaceint{V_{\rm ne}}{r} \left[ (\nabla \cdot \vect n)^2
   + (\nabla \times \vect n)^2 \right]   \\
    &=& \frac{K}{2}\spaceint{V_{\rm ne}}{r} \nabla n_{i}\cdot \nabla n_{i} + \nonumber \\
 & &   
  \frac{K}{2}\spaceint{V_{\rm ne}}{r} \nabla \cdot[ \vect n (\nabla \cdot \vect n) -(\vect n \cdot \nabla) \vect n] \;. \nonumber
\eea
$V_{\rm ne}$ denotes the volume occupied by the nematic film, $\vect n$ is the
director field ($\vect n^2 =1$), and the constant $K$ is of the order of $10^{-11}$ N
\cite{deG74}. 
The  total divergence term in the last line of Eq.~(\ref{eq:f_lc}) exhibits 
the so--called ``$K_{24}$--structure" and is unimportant for the bulk equations describing
the equilibrium configuration.
The surface free energies associated with the interfaces with air and substrate, 
respectively, (see Fig.~\ref{fig:nem_exp}) 
are described in terms of the Poulini expression \cite{Lav03}: 
\bea
 \label{eq:fs_lc}   
   {\cal F}^{\rm s}_{\rm ne} &=& \frac{W_1}{2} \int_{A_{\rm air-ne}}
  \!\!\!\!\!\!\!\!\!\!\!\! dA \;
   (\vect n \cdot \vect e_A)^2 + 
%  \nonumber \\ & &
   \frac{W_2}{2} 
   \int_{A_{\rm sub-ne} \bigcup A_{\rm coll-ne}}
  \!\!\!\!\!\!\!\!\!\!\!\!\!\!\!\!\!\!\!\!\!\!\!\!\!\!\!\!\!\!\!\!\!\!\!\!\! dA \;
   (\vect n \cdot \vect e_A)^2  \;.
\eea 
%{\tt New things included from here till Eq.(3)}
Here, $\vect e_A$ denotes the local surface normal unit vector pointing outwards from the
film or the colloid. Normal alignment is favored
for $W_i<0$ and parallel alignment for $W_i>0$. 
%
%and the director adopts the favored alignment over a length of the
%order of $K/|W_i|$.
%
Typically one has
$|W_i| \sim 10^{-5}$ N/m \cite{Lav03} so that the length scale $K/|W_i| \sim 1$ $\mu$m is smaller
than the range of droplet radii investigated in Ref.~\cite{Lav04}.
Thus, in the ``strong anchoring'' limit which we shall consider, %this case, 
the effect of the boundary terms is so strong that
as a first approximation it amounts to  
fixing the angle between the director and the surface normal. 
We shall adopt $W_1<0$ (normal alignment at the nematic--air interface) and $W_2>0$ 
(parallel alignment at the nematic--glycerol interfaces). Some consequences of 
deviations from the strong anchoring limit will be discussed in App.~\ref{sec:app1}.

These surface contributions to the free energy (``wetting energies'') 
are small corrections to the 
surface tensions which are mainly due to dispersion interactions.
We denote these non--nematic contributions to the surface tensions as  
$\gamma_1$ (colloid--air surface
tension), $\gamma_2$ (substrate--nematic surface tension) and $\gamma'$ (nematic--air
surface tension). Typically, these surface tensions are of the order of  $10^{-2}$ N/m.
Therefore, they are much larger than the constants $W_i$ which determine the nematic
contributions to the surface tension. We see that due to 
the above anchoring conditions the surface tension of the substrate--nematic interface
carries no nematic contributions due to $\vect n \cdot \vect e_A=0$ (Eq.~(\ref{eq:fs_lc})) 
whereas in the strong anchoring limit the full nematic--air surface tension 
is $\gamma= \gamma' + W_1/2 \approx \gamma'$ due to $\vect n \cdot \vect e_A=1$ 
(Eq.~(\ref{eq:fs_lc})).\footnote{
A genuine contribution to the surface tension of a nematic interface
arises if one takes into account the variation of the nematic tensorial order
parameter through the interface, described by, e.g., the Landau--de Gennes
free energy functional (generalizing Eq.~(\ref{eq:f_lc})). 
The magnitude of these contributions
can be estimated by the surface tension of the interface between the nematic and
the isotropic phase of a liquid crystal. Typically, such a surface tension
is also $O(10^{-5}$ N/m) $\sim |W_i|$ and therefore small compared to the dispersion
force contribution. Furthermore, a distorted director structure in the {\em bulk}
may also give rise to surface energy contributions on the boundaries. For an 
example see Ref.~\cite{deG74}, p. 131 and p. 174. Also in this case it can be
argued that the corresponding contributions to the surface tension do not exceed
$|W_i|$.  } 
%
%Ref.~\cite{Lav04} identifies the force associated to the anchoring
%``wetting energy'' $\propto W_2$ with the ultimate origin of the
%capillary attraction. We notice, however, that this argument is
%incomplete: if the particle is only partially wetted by the nematic
%phase (contact angle $\theta \in (0,\pi)$, as found experimentally),
%this force will be counteracted by the force associated to the
%``wetting energy'' of the particle--air interface. Therefore, the
%particle can be trapped at an undeformed fluid interface
%%without the need of a deformation of the meniscus 
%and remain in mechanical equilibrium \cite{Pie80} (this effect is
%described by the second term in the free energy functional in
%Eq.~(\ref{eq:F}) below). In the present case, however, there are also elastic
%forces pulling at the interface. Our goal is the computation of the
%interfacial deformation and the ensuing effective capillary
%interaction due to these elastic forces.

The canonical stress tensor $\pi_{ij}$ associated with the free energy 
expression in Eq.~(\ref{eq:f_lc})
is given by 
\bea
 \label{eq:stresstensor}
 \pi_{ij} = \frac{\partial{ f^{\rm b}}}{\partial n_{k,j}}\;n_{k,i} - \delta_{ij}\;f^{\rm b}
\eea
where $n_{k,i} =\partial n_k/\partial x_i$ (summation over $k$). 
The total stress tensor $\Pi_{ij}$ is obtained by
adding the contribution of the isotropic pressure $p$:
\bea
 \Pi_{ij} = \pi_{ij} - \delta_{ij}\;p\;.
\eea

\subsection{Macroscopically thick nematic film}
\label{sec:thick}

\begin{figure}[t]
 \begin{center}
  \epsfig{file=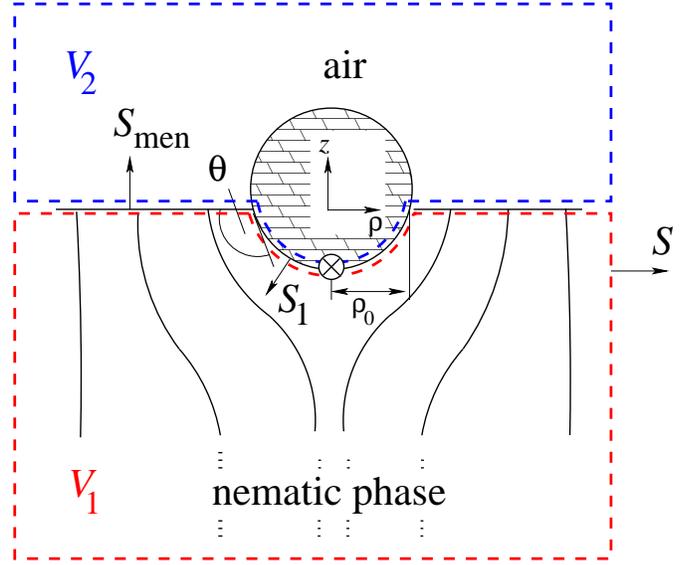, width=\columnwidth}
 \end{center}
  \caption{\label{fig:ref}
     In the reference configuration the whole system is divided into
      volumes $V_1$ and $V_2$. Volume $V_2$ (enclosed by the
      upper dashed curve) includes the air and the glycerol drop and volume $V_1$ (enclosed by
      the lower dashed curve) includes the nematic.
      The arrows indicate the
      direction in which the surfaces (including the infinitesimally
      displaced ones) are oriented: $S$
      encloses the whole system,  $S_{\rm men}$ is the interface
      between the nematic phase and air (acting as a meniscus), and $S_1$ is the 
      interface between the colloidal
      drop and the nematic phase. The director field and the topological 
      defect ($\otimes$) are
      indicated as in Fig.~\ref{fig:nem_exp}. The radius of the three--phase
      contact line is denoted by $\rho_0=R\sin\theta$ where $\theta$ is the contact angle
      of the air--nematic interface with the colloid of radius $R$. $\rho$ denotes
      the lateral distance from the vertical symmetry axis of the colloidal drop.}
\end{figure}

First we consider the limiting case $h \to \infty$ (i.e., very thick nematic films (see 
Fig.~\ref{fig:nem_exp})).\footnote{This was 
implicitly assumed also  by the authors of  Ref.~\cite{Lav04} in discussing Fig.~2 
therein.} Due to 
the small values of the elastic coupling constant, $K \ll \gamma\, R$, 
%(see Fig.~\ref{fig:nem_exp})
and
of the anchoring energy, $|W_i| \ll \gamma$, the equilibrium 
configuration of  a single colloid at the nematic--air interface deviates 
only slightly from
the reference configuration depicted in Fig.~\ref{fig:ref}. In this latter configuration, 
the interface
is flat and the colloid is positioned such that the contact angle fulfills Young's law
$\cos \theta = (\gamma_1-\gamma_2)/\gamma$.

The total force on the whole system reads (the
superscript $^{+(-)}$ denotes evaluation on the positive (negative)
side of the oriented surface, i.e., on the side the arrows in Fig.~\ref{fig:ref}
[which indicate the surface normals] point to (do not point to)):
\bea
 \oint_S  d{\bf A} \cdot {\mathbf \Pi} &=& \int_{V_1\bigcup V_2} 
  \!\!\!\!\!\! dV \; 
   (\nabla \cdot {\mathbf \Pi}) + \int_{S_{\rm men}\bigcup S_1} 
  \!\!\!\!\!\! d{\bf A} \cdot 
  ({\mathbf \Pi}^+ - {\mathbf \Pi}^-) \nonumber \\
  &=& - \int_{S_{\rm men}}  dA \; (\pi_{zz}+p_{\rm air}-p)  \, {\bf e}_z  + 
  \nonumber \\ &&
  \int_{S_1} d{\bf A}\cdot \left[{\boldsymbol \pi}+(p_{\rm air}-p){\bf 1}\right]
   \nonumber \\
  &=& - \int_{S_{\rm men}}  dA \; \pi_{zz}  \, {\bf e}_z  + 
  \int_{S_1} d{\bf A}\cdot {\boldsymbol \pi}  \; .
 \label{eq:Ftot}
\eea
In obtaining this equation we have applied Gauss' theorem. Furthermore
we have used the relation
$\nabla \cdot {\mathbf \Pi}=0$ in volumina $V_1$ and $V_2$ which is valid because 
the reference configuration is taken 
to be in force equilibrium. This also implies that the  isotropic pressures above the
interface ($p_{\rm air}$) and below it ($p$) are equal 
and that the director configuration is given by the 
corresponding Euler--Lagrange equilibrium equations following from the functional
in Eq.~(\ref{eq:f_lc}).
Since at the interface $S_1$ the colloidal drop is rigidly attached to the liquid crystal,
we can identify the vertical force $F$ on the colloid and the total force $F_\pi$
on the air--nematic interface by
\bea
 \label{eq:Fdef}
  F\;{\bf e}_z &=& \int_{S_1} d{\bf A}\cdot {\boldsymbol \pi}
\eea
and 
\bea 
 \label{eq:Fpi}
  F_\pi &=& -\int_{S_{\rm men}}  dA \; \pi_{zz} \; ,
\eea
respectively.
Mechanical isolation of the system means that the total force  
$\oint_S  d{\bf A} \cdot {\mathbf \Pi}$ acting  on it 
is zero which leads to
\bea
 \label{eq:f_eq_pi}
  F&=& F_\pi\;.
\eea 

For a given force on the colloid and a given stress on the interface, the
interface deformation relative to the reference configuration can be determined perturbatively.
To that end, we summarize briefly those results of Ref.~\cite{Oet05} which are pertinent
also for the present system. 
With the introduction of the  two small, dimensionless parameters 
\bea
  \label{eq:epsdef}
  \varepsilon_F = - F/(2\pi\,\rho_0\,\gamma) \quad \mbox{and} \quad
  \varepsilon_\pi = - F_\pi/(2\pi\,\rho_0\,\gamma)
\eea
one can  expand (up to second order in
$\varepsilon_F$ and $\varepsilon_\pi$) the free energy difference ${\cal F}$  
associated with the interface deformation $u(\rho\ge \rho_0)$ 
around a single colloid (see Fig.~\ref{fig:ref}) 
and with a vertical shift $\Delta h$ which is the difference of the  colloid center position
relative to that in  the reference configuration: 
\bea
  {\cal F} & \simeq & 
  2 \pi \gamma \int_{\rho_0}^{\infty} d\rho \; \rho \left[ \frac{1}{2}\left(\frac{d u}{d \rho} \right)^2 +
    \frac{u^2}{2 \lambda^2} + \frac{1}{\gamma} \pi_{zz} \, u \right] +
 \nonumber \\ &&
  \label{eq:F}
  \pi \gamma [ u(\rho_0) - \Delta h ]^2 -
  F \Delta h \;.
\eea
Here, $\lambda=(\gamma/(g\bar\rho_{\rm m}))^{1/2}$ is the capillary length associated 
with the interface where $g$ is the gravitational constant and $\bar\rho_{\rm m}$ is the
mass density of the nematic phase. This expression
for the free energy contains all surface free energy changes relative to the reference
configuration involving the interfaces between air, nematic,
or colloid. It also  contains the contributions due to  volume forces acting on the nematic 
(associated with $\lambda$)
and the energy change of the colloid upon vertical shifts (for further details see
Ref.~\cite{Oet05}).
Note that 
to leading (quadratic) order in $\varepsilon_\pi,\varepsilon_F$
the free energy change of the nematic due to the shifted interface
and due to a change in the director configuration with respect
to the reference configuration is captured 
by the term $\propto \int \pi_{zz}\,u$.
(The analogous textbook argument for electrostatics \cite{Sch98} can be 
easily generalized
to the nematic case described by the free energy expression in Eq.~(\ref{eq:f_lc}).) 

Minimizing ${\cal F}$ with respect to $u(\rho)$ and $\Delta h$ and 
focussing on  the regime $\rho \ll \lambda$ yields 
\bea
  \label{eq:sol1}
  u (\rho) \simeq \rho_0 (\varepsilon_\pi - \varepsilon_F) \ln\frac{C \lambda}{\rho}
  + \frac{1}{\gamma} \int_{\rho}^{+\infty} \!\!\!\! d\sigma \; \sigma \, \pi_{zz}(\sigma) \ln \frac{\sigma}{\rho} \;, \quad
\eea
with $C \simeq 1.12$. We see that in the case of an isolated system
$(\varepsilon_\pi= \varepsilon_F)$ the logarithmic part of $u(\rho)$ vanishes. The 
second term on the rhs of Eq.~(\ref{eq:sol1}) leads to 
$u(\rho \to \infty) \propto \rho^{-n+2}$
if $\pi_{zz} \propto \rho^{-n}$ and thus describes a shorter--ranged power--law
decay of the interface deformation. 

The absence of logarithmic deformations for an isolated system has been derived here
under certain simplifying conditions (small interfacial deformation everywhere, rotational symmetry).
%which allows using the reference configuration. 
In App.~\ref{sec:app2} we demonstrate that this conclusion
holds in general.
% less restrictive case that the interface deformation is small
% only far away from the particles.

\subsubsection{Asymptotic director configuration and elastic force between colloids}

%Next we derive
%the exponent $n$ of the asymptotic stress field $\pi_{zz}$ for our simple 
%director model, defined by Eq.~(\ref{eq:f_lc}) and 
%perpendicular anchoring at the interface as realized experimentally in 
%Ref.~\cite{Lav04}. 
In Refs.~\cite{Pou97,Lub98} it has been shown that a colloidal drop 
immersed in the bulk of a liquid crystal
is accompanied by a single counterdefect such that the total topological charge is zero
(here, the volume occupied by the colloid contains a topological charge which
may be represented by a virtual defect inside the colloid) and
the asymptotic behavior of the director field is of dipolar character.
Based on similar considerations we shall show that for a colloidal drop located at the
air--nematic interface the boundary conditions for that interface impose a 
quadrupole--like  asymptotic 
behavior of the director field. Macroscopically far from the colloid the director
is oriented parallel to the $z$ axis.
Accordingly, at large but finite distances $r$ the director is
given by $\vect n(\vect r) \simeq (n_1,n_2,1-O(n_1^2,n_2^2))$ and the bulk
free energy corresponding to Eq.~(\ref{eq:f_lc}) is  given by
\bea
 \label{eq:fasy}
  {\cal F}^{\rm b}_{\rm ne} \simeq \frac{K}{2} \spaceint{V_{\rm ne}}{r} 
    \left( \sum_{i=1,2} \left( \nabla n_i 
    \right)^2 + O (n_i^4) \right)\;.
\eea
Here we have discarded the total divergence term in the free energy expression
(\ref{eq:f_lc}). It is unimportant for the bulk equations and it adds a mere
constant to the free energy because the director is anchored normally at 
the boundary (the nematic--air interface).
Thus for each component
$i=1,2$ the equilibrium director field fulfills the Laplace equation
\bea
 \label{eq:lapl}
   \Delta n_i = 0\;.
\eea
Analogously to electrostatics, the asymptotic solution for $n_i$ can be expanded
in terms of multipoles. To this end we  consider the reference configuration in
 Fig.~\ref{fig:ref}. 
We choose as the origin of the coordinate system the center 
of the circle formed by the planar three--phase contact line. 
The solution for the director field has to fulfill the following requirements:
($i$) rotational covariance around the $z$-axis\footnote{If $\bf D$ specifies 
the transformation matrix for such a rotation then this
requirement is given by $\vect n (\bf D\cdot \vect r) = \bf D \cdot \vect n
(\vect r)$.} and ($ii$) $n_i(x,y,z=0)=0$.
Analyzing the multipole {\em ansatz} (with $\vect r=(r_1, r_2, r_3)$)
%\alpha,\;\alpha=1,2,3)$)
\bea
 \label{eq:es_mul}
  n_i &=& q_i\,\frac{1}{r} + \sum_{\alpha=1}^3 P_{i\alpha}\,\frac{r_\alpha}{r^3} +  
      \sum_{\alpha,\beta=1}^3 Q_{i\alpha\beta}\,\frac{r_\alpha\,r_\beta}{r^5}
 + \dots
\eea
it follows that rotational covariance requires $q_i=0$, 
$ P_{i\alpha} = P\,\delta_{i\alpha} + P_{\rm mag}\,\epsilon_{i\alpha 3}$ 
and $Q_{i\alpha\beta}=Q'_\beta\,\delta_{i\alpha} + 
Q'_{{\rm mag},\beta}\,\epsilon_{i\alpha 3}$ ($\epsilon_{ijk}$ is the Levi-Civit\`a tensor).
$P_{\rm mag}$ and $Q'_{{\rm mag},\beta}$ are dipole and quadrupole moments,
respectively, for a director field of ``magnetic" type, i.e., for
which ${\rm div}\,n_i=0$ holds. 
The boundary condition ($ii$) at the interface with the air further imposes 
$P=P_{\rm mag}=0$ and
$Q'_\beta= Q\, \delta_{\beta3}$, $Q'_{{\rm mag},\beta}= Q_{\rm mag}\, 
\delta_{\beta3}$. It appears to be difficult geometrically to match
the asymptotic solution of ``magnetic" type  with a solution near the
colloid which obeys parallel anchoring at the colloid surface. 
%\textcolor{blue}{\sout{Furthermore,
%if we assumed weak anchoring ($|W|R/K \ll 1$), the magnetic
%quadrupole would vanish identically as can be shown in a perturbative
%calculation along the lines of Ref.~\cite{Ruh97}. }}
Therefore we 
discard the magnetic quadrupole,
 i.e., the leading asymptotic term is given by
the remaining, ``electric" quadrupole term:
\bea
   \label{eq:quadrupole}
  n_i  &=& Q\; \frac{z\,r_i}{r^5} +\dots \qquad (z \equiv r_3) \;. 
\eea 
This is at variance with Ref.~\cite{Lav04} 
(see Eq.~(3) therein and the considerations in the paragraph above that equation 
which assume
a dipole field) but it is consistent with the analysis in Ref.~\cite{Lub98}. 
Dimensional analysis yields $Q=O(R^3)$ \cite{Lub98}. 
Note that we have derived the asymptotic behavior of the director field at the
interface using strong anchoring at the interface ($n_1=n_2=0$). 
In App.~\ref{sec:app1}
we discuss corrections to strong anchoring which are, however, of subleading
character and leave the leading behavior (Eq.~(\ref{eq:quadrupole})) unchanged.

The asymptotic elastic interaction between two colloids in the bulk at distance $d$ accompanied
by a quadrupolar director deformation has been analyzed in Ref.~\cite{Ruh97}
(for weak anchoring) and in Refs.~\cite{Ram96,Lub98} (using a coarse--graining
method, applicable also for strong anchoring) yielding identical results. For the present configuration (distance
vector perpendicular to the asymptotic director) the elastic
potential is repulsive  and varies as
\begin{equation}
  V_{\rm el} \propto \frac{K\,Q^2}{d^5} \propto \gamma \rho_0^2\,
   \varepsilon_F\left(\frac{\rho_0}{d}\right)^5\;. 
\end{equation}
We have used that the dimensionless force parameter $\varepsilon_F$ is proportional to
$Q^2$
which actually follows from Eqs.~(\ref{eq:Fdef}), (\ref{eq:f_eq_pi}), (\ref{eq:epsdef}), and  
(\ref{eq:pizz_single}) below.
Note that we have simply extrapolated the {\em bulk} results for two interacting
colloids which cause asymptotically quadrupolar deformations of the director.
 This appears
to be reasonable because the asymptotic director field in the nematic phase for 
the interface
problem is  precisely that of the bulk solution in the lower half plane
and because the bulk solution is antisymmetric with respect to $z \to -z$, thus
respecting the boundary condition $n_i(x,y,z=0)=0$. However, the precise
numerical value of the quadrupole moment $Q$ might be rather different for the case
of the colloid trapped at the interface as compared to the bulk case.

\subsubsection{Asymptotic behavior of the stress on the interface and meniscus--induced 
effective potential between colloids}

The asymptotic behavior of the stress tensor component $\pi_{zz}$ at the interface
follows from inserting 
Eq.~(\ref{eq:quadrupole}) into Eq.~(\ref{eq:stresstensor}):
\bea
 \left. \pi_{zz}\right|_{\rm interface} &=&
%  \left.\frac{K}{2} \sum_{i=1}^2 \left(n_{i,z}n_{i,z}-n_{i,r_1}n_{i,r_1}-n_{i,r_2}n_{i,r_2}
  \left.\frac{K}{2} \sum_{i=1}^2 \left(n_{i,z}^2-n_{i,r_1}^2-n_{i,r_2}^2
   \right)\right|_{z=0} 
 \nonumber \\  
 \label{eq:pizz_single}
  &\stackrel{r\to\infty}{\longrightarrow} &  \frac{K}{2}\,Q^2 \,\frac{1}{\rho^8} \;, 
\eea
($\rho^2=r_1^2+r_2^2$).
Consequently the interface
deformation around a single colloid for a mechanically isolated system obeys
$u(\rho\to\infty)\propto \rho^{-6}$ (see Eq.~(\ref{eq:sol1})). 

For the problem of two identical colloids  located at ${\boldsymbol \rho}_1$ and 
${\boldsymbol \rho}_2$ (vectors are defined in the interface plane $z=0$) a distance
$d=|{\boldsymbol \rho}_1-{\boldsymbol \rho}_2| \gg R$ apart the expression for the
free energy is a straightforward generalization of the one for the single--colloid
free energy given in Eq.~(\ref{eq:F}):
\bea
  \label{eq:F2}
  {\hat {\cal F}} &= &
  \gamma \int_{S_{\rm men}} \!\!\!\!\!\! d^2 \rho \;
  \left[  \frac{|\nabla \hat{u}|^2}{2} + \frac{\hat{u}^2}{2 \lambda^2} -
    \frac{\hat{\pi}_{zz}}{\gamma} \, \hat{u} \right] +
 \\ \nonumber & &
   \sum_{i=1,2} \left\{
    \frac{\gamma}{2 \rho_{0}} \oint_{\partial S_i} \!\!\! d\ell \; [\Delta \hat{h}_i - \hat{u}]^2
    - \hat{F}_i \Delta \hat{h}_i \right\} .
% \nonumber
\eea
Here, $\hat{F}_i$ denotes the force on colloid $i$ and  $\Delta \hat{h}_i$ is the relative
position of its center. The integration domain $S_{\rm men}$ is the whole interface plane
except for the two circular disks bordered by the (reference configuration) contact lines  
$\partial S_i$. The meniscus--induced effective potential 
is the difference between the equilibrium free energy of the two colloids at distance $d$
and their free energy at macroscopic distance:
\bea
  V_{\rm men}(d) = {\hat {\cal F}}_{\rm eq} (d) - {\hat {\cal F}}_{\rm eq} (d\to\infty) \;.
\eea
As before, minimization with respect to
$\hat u(\boldsymbol \rho)$ and $\Delta \hat h_i$ renders the equilibrium free energy.

The behavior of $V_{\rm men}(d)$ has been analyzed  in detail in 
Refs.~\cite{Oet05,Dom07}. Here we summarize these results as far as they are relevant
for the present problem.
The interfacial stress $\hat \pi_{zz}$ % and the interface deformation field $\hat u$ 
may be decomposed generally as
\bea
  \hat \pi_{zz}(\boldsymbol \rho)& = &\pi_{zz}(|\boldsymbol \rho-\boldsymbol \rho_1|) + \pi_{zz}(|\boldsymbol \rho-\boldsymbol \rho_2|) 
  + 2\,\pi_{zz, {\rm m}}(\boldsymbol \rho)
 \nonumber \\
  \label{eq:pihat}
 & \equiv & \pi_{zz,1}+\pi_{zz,2}+2\,\pi_{zz, {\rm m}} \;.% \\
%  \label{eq:uhat}
%  \hat u(\boldsymbol \rho)& = &u(|\boldsymbol \rho-\boldsymbol \rho_1|) + u(|\boldsymbol \rho-\boldsymbol \rho_2|) + u_{\rm m}(\boldsymbol \rho)
%  \equiv u_1+u_2+u_{\rm m} \;. 
\eea
Here, $\pi_{zz, i}$ denotes the stress %and  interface deformation field
around  colloid $i$ which pertains to the problem of a single colloid. 
To quadratic order the asymptotic director field around 
two colloids is given by the superposition of the components $n_i$ of the
single--colloid solutions and thus to this order we recover the decomposition
in Eq.~(\ref{eq:pihat}) with the mixed
component of stress field $\pi_{zz, {\rm m}}$ given by
\bea
  \pi_{zz, {\rm m}} =
 \frac{K}{2}\;Q^2\; \frac{(\boldsymbol \rho-\boldsymbol \rho_1) \cdot (\boldsymbol \rho-\boldsymbol \rho_2)}
{|\boldsymbol \rho-\boldsymbol \rho_1|^5\; |\boldsymbol \rho-\boldsymbol \rho_2|^5} \;.
\eea

It turns out that  
for a system under an external force $(\varepsilon_\pi \not = \varepsilon_F)$
the mixed term $\pi_{zz, {\rm m}}$ does not contribute to the leading
term in $V_{\rm men}$. %and $u_{\rm m}$. 
Thus in the case of a non--vanishing external force this leading
contribution to $V_{\rm men}$ is obtained by a superposition 
{\em ansatz} which  consists in 
approximating the interfacial deformation and the total %deformation and 
stress field by the sum of the respective single--colloid quantities only
($\hat \pi_{zz}\approx \pi_{zz,1}+\pi_{zz,2}$,
 $\hat u \approx u_1+u_2$) \cite{Oet05,Dom07}:
\bea
 \label{eq:vmen_noniso}
 V_{\rm men}(\rho_0 \ll d \ll \lambda) &\simeq&   - 2\pi\;\gamma\; \rho_0^2\,
   (\varepsilon_\pi - \varepsilon_F)^2 \, \ln \frac{C\lambda}{d}  \qquad
  \\ && \nonumber
     (\varepsilon_\pi \not = \varepsilon_F)\;.
\eea
For an isolated system  $(\varepsilon_\pi= \varepsilon_F)$ and for the stress
given in Eq.~(\ref{eq:pizz_single}) the superposition approximation 
$\hat \pi_{zz}\approx \pi_{zz,1}+\pi_{zz,2}$
yields $V_{\rm men} \propto \varepsilon_F^2/d^8$ as the dominant term. 
However, this is not the leading term, which rather stems from 
$\pi_{zz,{\rm m}}$. This term has two peaks around the colloid centers and 
therefore close to the colloids it can be approximated by
\bea
  \label{eq:pimmpeak}
  \pi_{zz,{\rm m}} \approx \frac{K \,Q^2}{2\,d^4} \sum_{i=1}^2 (-1)^i
  \frac{ \vect e_d \cdot(\boldsymbol \rho-\boldsymbol \rho_i)}
  {|\boldsymbol \rho-\boldsymbol \rho_i|^5} 
\eea
where $\vect e_d = (\boldsymbol \rho_2-\boldsymbol \rho_1)/d$.
%This mixed stress gives rise to mixed contributions $u_{\rm m}$ which are also
%centered around the colloids and whose overall amplitude is determined by the 
%prefactor of $\pi_{zz,{\rm m}}$ and thus we have $u_{\rm m} \propto d^{-4}$.
As discussed in Ref.~\cite{Dom07}, the qualitative behavior of  $V_{\rm men}(d)$  
is captured by the integral over $\pi_{zz,{\rm m}}$:
\bea
 \label{eq:vmen_iso}
  V_{\rm men}(d) &\propto & \gamma\rho_0^3\,\varepsilon_F \int_{S_{\rm men}} \!\!\! d^2\rho \, 
   \pi_{zz,{\rm m}}(\boldsymbol \rho) 
  \propto  \gamma\rho_0^2\,\varepsilon_F^2\,\left(\frac{\rho_0}{d}\right)^5 \qquad
 \\ && \nonumber
  (\varepsilon_\pi = \varepsilon_F)\;.
\eea
We note that this form of the mixed stress and thus the leading behavior of 
$V_{\rm men}(d)$
is formally analogous to the contributions of the
electric field components parallel to the interface in the case of charged colloids
\cite{Dom07}.  
The meniscus--induced potential $V_{\rm men}$ is repulsive and 
falls off asymptotically\footnote{Superficially one would expect a 
leading decay 
$V_{\rm men}\propto d^{-4}$ as displayed in Eq.~(\ref{eq:pimmpeak}). However, due to
the geometric factor in the numerator of Eq.~(\ref{eq:pimmpeak}), this apparent
leading order vanishes upon integration.} $\propto
d^{-5}$, as does the likewise repulsive elastic 
potential $V_{\rm el}$. We note that 
$V_{\rm el} \propto |\varepsilon_F|$ and $V_{\rm men} \propto \varepsilon_F^2$, so that
the meniscus--induced potential is small compared to the elastic one for
the parameters of the experiment reported in Ref.~\cite{Lav04}.

In order to explain the experimentally observed attractions, in Ref.~\cite{Lav04} 
a perturbative 
picture similar to the one presented
above was suggested, but the effect of the interface stress $\pi_{zz}$ was
neglected completely. (A similar error has been made in Ref.~\cite{Nik02} which
the authors of Ref.~\cite{Lav04} refer to.) 
%{\tt The following sentence has been rewritten:} 
In a heuristic way, only an ``upward" force on the colloid (perpendicular to the interface)
 associated with
the anchoring ``wetting" energy at the nematic--particle interface
has been invoked in Ref.~\cite{Lav04}, neglecting
the force on the interface described by $\pi_{zz}$.
% which, as we have seen, is incomplete.
% a change of the director
% field between the reference and the deformed configuration which can be seen
% to be of higher order
% in the parameters $\varepsilon_\pi, \varepsilon_F$. 
In this way, the unbalance of the force
gives rise to a logarithmic term in the meniscus
deformation and in the meniscus--induced potential. However, mechanical isolation 
(i.e., force balance) renders
the meniscus--induced potential actually repulsive and shorter--ranged 
(see Eq.~(\ref{eq:vmen_iso})), apparently in contrast to the experimental results.
A logarithmically varying potential can only arise if mechanical isolation is violated (see App.~\ref{sec:app2}). 
Below we shall investigate whether a net force on the system ``colloid and interface"
may appear if the
thickness of the nematic phase is finite, as it is the case in the actual experiment.

\subsection{Finite thickness of the nematic film}
\label{sec:thin}

In our discussion of a finite film thickness of the nematic phase we shall consider two cases:
\begin{enumerate}
\item  The anchoring of the nematic director at the surface of the bottom substrate is 
perpendicular as it is the case at the upper interface with the air. 
This case bears a strong formal resemblance to  charged
colloids on water surfaces which have been  discussed in Refs.~\cite{Oet05,Dom07}. 
\item The anchoring at the bottom substrate surface is parallel (as in the experiment reported in
Ref.~\cite{Lav04}). At large lateral distances from the colloids, this leads to a director
field which gradually rotates from parallel orientation at the bottom substrate to the
perpendicular orientation at the upper interface. 
\end{enumerate}
%{\tt new intermediate summary}
%In combination with the normalization
%condition ${\bf n}^2=1$ this leads to a strong net force
%and consequently to a rather strong interface deformation. This effect is peculiar 
%to this experimental system.  

For both cases the total force on the system -- comprising air, nematic film, 
colloid, and the substrate -- must be zero. This leads to
(see Fig.~\ref{fig:ref1}(a) and compare with Eq.~(\ref{eq:Ftot}))
\bea
 0 &=& \oint_S  d{\bf A} \cdot {\mathbf \Pi} \nonumber \\
   &=& \int_{V_1\bigcup V_2\bigcup V_2'} \!\!\!\!\!\!\!\!\! dV \; 
   (\nabla \cdot {\mathbf \Pi}) + \int_{S_{\rm men}\bigcup S_1\bigcup S_{\rm sub}} \!\!\!\!\!\!\!\!\! d{\bf A} \cdot 
  ({\mathbf \Pi}^+ - {\mathbf \Pi}^-) \nonumber \\
  &=& - \int_{S_{\rm men}}  dA \; (\pi_{zz}+p_{\rm air}-p)  \, {\bf e}_z  +
 \label{eq:Ftot1} \\
 && \int_{S_1} \!\!\!d{\bf A}\cdot \left[{\boldsymbol \pi}+(p_{\rm air}-p){\bf 1}\right]
  + \int_{S_{\rm sub}} \!\!\!\!\!\!dA \; (\pi_{zz}+p_{\rm sub}-p)  \, {\bf e}_z  \; .
 \nonumber
\eea
For reasons which will become clear in the  discussion of  the second anchoring case, 
we consider the isotropic pressures in the substrate and air, $p_{\rm sub}$ and
$p_{\rm air}$, respectively, not necessarily to be equal to the pressure $p$ in the 
nematic film. The second equation in Eq.~(\ref{eq:Ftot1}) follows from the equilibrium
condition $\nabla \cdot {\mathbf \Pi}=0$ which holds  in all volumina.

\subsubsection{Perpendicular anchoring at both interfaces}

\label{sec:aligned}

\begin{figure}[t]
 \begin{center}
  \epsfig{file=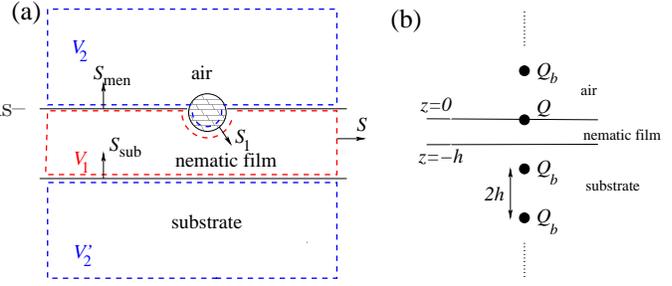, width=\columnwidth}
 \end{center}
  \caption{\label{fig:ref1}
    The reference configuration for a nematic film (a) differs from the reference 
    configuration
    shown in Fig.~\ref{fig:ref} by the addition of the substrate volume $V_2'$.
    The interface between the  substrate and the nematic film is denoted by 
    $S_{\rm sub}$.
    Panel (b) shows the string of image quadrupoles $ Q_b=Q$ which are needed 
    to fulfill the boundary conditions for perpendicular anchoring of the director 
    field (originating
    from the colloid quadrupole $Q$) at both interfaces confining the nematic film. 
    The distance between any two nearest neighbor quadrupoles is $2h$.
   }
\end{figure}

As discussed above, the presence of the  colloid asymptotically generates a 
quadrupolar director field  
which fulfills the boundary condition at the nematic--air interface.
In order to fulfill the
boundary condition at the substrate--nematic interface, an image quadrupole
of the same strength $Q$ is needed which, however, leads to a violation of the
nematic--air interface boundary condition and requires a second image quadrupole etc.
Continuation of this process leads to a string of image quadrupoles as 
depicted in Fig.~\ref{fig:ref1}(b). This string of quadrupoles
generates a stress field ${\boldsymbol \pi}$ which vanishes for large lateral distances.
Therefore the isotropic pressures must be equal in all volumina:
$p_{\rm sub}=p_{\rm air}=p$. From Eq.~(\ref{eq:Ftot1}) one finds that the difference
between the force $F$ on the colloid and the integrated stress 
$F_\pi$ at the nematic--air interface,
\bea
 \label{eq:df1}
  F- F_\pi &=& \Delta F = - \int_{S_{\rm sub}}  dA \; \pi_{zz}\;,
\eea
is given by the integrated stress over the substrate surface, i.e., the total force
on all quadrupoles above the substrate surface exerted by the image quadrupoles
in the substrate. Expressed in terms of the force $F_{Q-Q}$ between two quadrupoles
at distance $2h$, we find
\bea
 -\Delta F / F_{Q-Q} & =&   \sum_{n=1}^\infty \frac{1}{n^6} + \sum_{n=2}^\infty 
 \frac{1}{n^6} + \dots \nonumber \\ 
    &  = &  \sum_{n=1}^\infty \frac{1}{n^5} = \zeta(5) \approx 1.04  \;.
\eea
In this equation, the first sum is the total force (divided by $F_{Q-Q}$) on the first
quadrupole above the substrate exerted by all quadrupoles in the substrate,
the second sum is the total force (divided by $F_{Q-Q}$) on the second quadrupole
above the substrate etc. 
For the force between two quadrupoles at distance $2h$ we find, 
using Eqs.~(\ref{eq:quadrupole}),
(\ref{eq:stresstensor}), and (\ref{eq:df1}):
\bea
 \label{eq:fqq}
   F_{Q-Q} & =& \frac{5}{6}\pi\,K\,\frac{Q^2}{h^6}\;.
\eea
Thus mechanical isolation for the system ``colloid and \linebreak nematic--air interface" 
is violated and the total force $\Delta F$ on this system 
is (up to the factor 1.04) given by the quadrupole force $F_{Q-Q}$.
Nevertheless the magnitude of the corresponding induced logarithmic capillary potential
(see Eq.~(\ref{eq:vmen_noniso}) with $\varepsilon_\pi - \varepsilon_F = \Delta F / (2\pi\gamma\rho_0)$) is small, because $Q \sim R^3$ due to dimensional arguments
\cite{Lub98}.
For parameters similar
to the ones appearing in Ref.~\cite{Lav04} 
($R/h\simeq 10^{-1}$, $K/(\gamma R)\simeq 10^{-4}$)
we find $V_{\rm men} \simeq 10^{-14}\, k_B T\;\ln(R/d)$, which is 
unimportant for the actual intercolloidal interaction. 
%
%{\tt I have corrected the numerical estimate of the amplitude, which
%  is now much smaller than quoted before. I have done the same in the
%  next subsection, and in the Conclusions with the power-law
%  dependence on $R/h$. It seems that you inadvertendly quoted the
%  amplitude of the \mbox{{\em deformation} ($\propto \Delta F$)}
%  whenever you really meant the amplitude of the \mbox{{\em
%      interaction energy} ($\propto (\Delta F)^2$)}. Please check.}
 
The elastic potential $V_{\rm el}(d)$ between two colloids 
is the interaction between the second quadrupole 
and the first quadrupole together with its string of image quadrupoles.
Using the solution given in Refs.~\cite{Ram96,Ruh97} we have checked
that $V_{\rm el}(d)$ remains repulsive.  For $d < h$ the overall magnitude of $V_{\rm el}(d)$
is somewhat weakened,
whereas for $d \gg h$ a crossover to $V_{\rm el}(d) \propto \exp(-d/h)$ is observed\footnote{This result can be obtained more easily by solving the field equations 
$\Delta n_i=0$ in cylindrical coordinates rather than by using the image quadrupoles.}.

\subsubsection{Parallel anchoring at the bottom substrate }

We assume that the substrate induces a preferred in--plane axis for the director 
orientation which we take to be the  $x$--axis. With no colloid present at the nematic--air 
interface, the equilibrium director field is given by
\bea
 \label{eq:efield}
  \vect n_0=  \begin{pmatrix}  \sin(-q_0 z) \\ 0 \\ \cos(-q_0 z) \end{pmatrix}
  \;, \qquad q_0 = \pi /(2 h) \;,
\eea 
with the consequence that both at the nematic--air and at the nematic--substrate 
interface a constant stress is acting:
\bea
  \pi_{0,zz} = \frac{K}{2}\,q_0^2\;.
\eea
For the unperturbed interface to be in equilibrium, the air and substrate pressures
differ from the pressure in the liquid crystal: $p-p_{\rm sub\,[air]}=
\pi_{0,zz}$. 

We now introduce a single colloid at the nematic--air interface in the reference 
configuration. 
Due to this pressure difference, the force on the colloid
and the integrated stress over the nematic--air interface are
given  by (see Eq.~(\ref{eq:Ftot1}))
\bea
 \label{eq:f1}
  F\;{\bf e}_z &=& \int_{S_1} d{\bf A}\cdot ({\boldsymbol \pi}-\pi_{0,zz}\,{\bf 1})\;, \\
 \label{eq:fpi1}
  F_\pi &=& \int_{S_{\rm men}}  dA \; (\pi_{zz}- \pi_{0,zz}) \; ,
\eea
and the total excess force on the system ``colloid  and nematic--air interface"
is determined by
\bea
 \label{eq:df2}
  F- F_\pi &=& \Delta F = - \int_{S_{\rm sub}}  dA \; (\pi_{zz}-\pi_{0,zz})\;.
\eea
In order to calculate the director field $\vect n$ and the stress tensor ${\boldsymbol\pi}$
in the presence of the colloid, we introduce the auxiliary director deformation 
fields $v(x,y,z)$ and $w(x,y,z)$ which parametrize the deviations from the unperturbed 
director field $\vect n_0$ and which are small at large distances from the colloid:
\bea
 \label{eq:nlin}
  \vect n &=& \begin{pmatrix} \sin(-q_0 z + v)\cos w \\ \sin w \\ \cos(-q_0 z + v)\cos w
            \end{pmatrix} \approx \vect n_0 + \\
 \nonumber  & &
   \begin{pmatrix}  \cos(q_0 z)\;v + \frac{1}{2}\sin(q_0 z)(v^2+w^2) \\ w \\ \sin(q_0 z) \;v -\frac{1}{2}\cos(q_0 z)(v^2+w^2)  \end{pmatrix}
  + O((v,w)^3) \;.\quad
\eea
The first equality in Eq.~(\ref{eq:nlin}) is a general parametrization of
the director field $\vect n$ in terms of the auxiliary fields $v,w$ 
which fulfills $\vect n^2=1$.
These auxiliary fields are taken to be zero at the substrate and the nematic--air interface,
i.e., the boundary conditions are $v(x,y,0)=v(x,y,-h)=w(x,y,0)=w(x,y,-h)=0$.
The nematic free energy of the film up to order $O(v^2,w^2)$ is obtained
by inserting Eq.~(\ref{eq:nlin}) into Eq.~(\ref{eq:f_lc}) for the Frank  free energy
after dropping the total divergence of the $K_{24}$--type.
Using the boundary conditions for $v$ and $w$ we obtain \cite{Fuk02}
(with the notation introduced in Eq.~(\ref{eq:stresstensor})):
\bea
  {\cal F}^{\rm film}_{\rm ne} = {\cal F}_0 + \frac{K}{2}\int_{V_{\rm film}} d^3r 
    \left( v_{,i}^2 + w_{,i}^2 - q_0^2\;w^2 \right)\;.
\eea
Here, ${\cal F}_0$ is the free energy of the film without colloid. At first
glance it is not evident  that
this free energy is positive definite. However, the boundary conditions
on $v$ and $w$ ensure positivity \cite{Fuk02}. Upon minimization we find
$\Delta v=0$, i.e., the deformation field  $v$ of the director 
can again be expanded in terms of electrostatic multipoles (Eq.~(\ref{eq:es_mul})).
On the other hand, the solution for $w$ must fulfill the Helmholtz equation
$(\Delta + q_0^2)w=0$ and can be expanded in terms of multipoles as follows:
\bea
 \label{eq:hel_mul}
  w(r,\theta,\phi)& =& \frac{1}{\sqrt{q_0 r}}\; \sum_{j=0}^\infty\sum_{m=-j}^{j} 
  Y_{jm}(\theta,\phi)\; \\
  \nonumber  & & \left( w^J_{jm}\; J_{j+1/2}(q_0 r) + w^Y_{jm}\; Y_{j+1/2}(q_0 r) \right)\;
    ,
\eea
where $J[Y]_{j+1/2}(r)$ are the spherical Bessel functions of the first
[second] kind and $Y_{jm}(\theta,\phi)$ 
are the usual spherical harmonics for the standard set of spherical coordinates
$r,\theta,\phi$. 
(The origin is again taken as the center of the circular three--phase contact line.)
The coefficients $w^{J[Y]}_{jm}$ are dimensionless multipole moments.

The Dirichlet boundary conditions for $v$ and $w$ at the two interfaces can be fulfilled
as before by constructing the full solution in terms of multipoles around the colloid
and corresponding image multipoles as shown in Fig.~\ref{fig:ref1}. 
Since rotational covariance is broken by the parallel substrate anchoring,
the solution for $v$ contains a nonzero dipole contribution.
Nevertheless, the director field
still obeys a reflection symmetry
with respect to the $xz$--plane: $v(x,y,z)=v(x,-y,z)$ and
$w(x,y,z)=-w(x,-y,z)$. Therefore, the
leading asymptotic behavior for $v$  is given by
\bea
 \label{eq:uasy}
  v = P_v \frac{z}{r^3} + Q_v \frac{zx}{r^5} +\dots + v_{\rm image} \;.
\eea
The dipole contribution should vanish for $h\to \infty$. If one
assumes a power--law dependence on $h$, dimensional analysis for 
$P_v$ leads to 
\bea
 \label{eq:kappadef}
 P_v = O(R^2\;(R/h)^\kappa) \qquad (\mbox{with}\; \kappa>0)\;.
\eea
The precise functional form of $P_v$  turns out to be unimportant for 
the subsequent calculations.
We note that an asymptotic solution with a nonvanishing
$x$--component of the dipole moment cannot appear because it would not fulfill
the boundary conditions and the reflection symmetry $w(x,y,z)=-w(x,-y,z)$ excludes
any dipolar contribution in the solution for $w$.\footnote{
In this respect the pictorial argument given in Ref.~\cite{Lav04} (see Fig.~3(b) therein)
is slightly misleading (at least in an asymptotic sense): there the
 assumed director configuration around the colloids for the
case of a finite film thickness is a tilted dipole with a nonvanishing component
$v \propto P_{v,x} x/z^3$. 
Actually, the broken symmetry in $x$--direction rather
enters through a tilted quadrupole.}  Therefore the leading asymptotic
behavior of $w$ takes the form
\bea
  w &=&  \frac{zy}{r^2}\;\frac{1}{\sqrt{q_0 r}} \left( Q^J_w\;J_{5/2}(q_0 r) +
   Q^Y_w \;Y_{5/2}(q_0 r) \right) + \dots 
  \nonumber \\&& + w_{\rm image} \;.
 \label{eq:vasy1}
\eea
We note that there are two quadrupolar contributions (with dimensionless moments
$Q^J_w$ and $Q^Y_w$)
 which show an oscillatory behavior 
for radial distances $r \gg h$ (i.e., $q_0 r \gg 1$). On the other hand, for 
large film thicknesses
there is an intermediate asymptotic regime $R \ll r \ll h$:
\bea
 \label{eq:vasy2}
  w (q_0 R \ll q_0 r \ll 1) &=& \frac{zy}{r^2}\;\sqrt{\frac{2}{\pi}} \left(
    Q^J_w\;\frac{(q_0 r)^2}{15} -   Q^Y_w \; \frac{3}{(q_0 r)^3} \right)
 \nonumber \\
  && + \dots + 
   w_{\rm image} \;.
\eea
For large film thicknesses the solution for the director field
near the nematic--air interface in the regime $R \ll r \ll h$ should coincide
with the solution for macroscopically thick films (Eq.~(\ref{eq:quadrupole})).
Near the nematic--air interface, one has $v\approx n_x$, $w\approx n_y$. Therefore
one recovers the rotationally covariant quadrupole solution of Eq.~(\ref{eq:quadrupole})
for  $Q_v=Q$ (obtained by equating Eq.~(\ref{eq:uasy}) with Eq.~(\ref{eq:quadrupole})
for $i=1$, i.e., $r_1=x$)  and $-3\sqrt{\di 2/\pi}\,Q^Y_w/q_0^3=Q$ (obtained by equating 
Eq.~(\ref{eq:uasy}) with Eq.~(\ref{eq:quadrupole}) for $i=2$, i.e., $r_2=y$).
Since $Q^Y_w \sim q_0^3 Q$ and $Q = O(R^3)$, $Q^Y_w=O([q_0 R]^3)$ is
a very small number.

The Dirichlet boundary conditions for $w$ at the substrate and at the 
nematic--air interface
enforce that the contribution to $w$ due to the quadrupole $Q^J_w$ and all corresponding image
quadrupoles is zero. This holds also for the contribution of  all higher multipoles of 
degree $j$ 
for which the  radial dependence is given by $J_{j+1/2}(r)$. 
The only solution of the Helmholtz equation
which fulfills the boundary conditions and which, as an additional requirement, 
is smooth everywhere,
is $w\equiv 0$. Since the Bessel functions of the first kind are smooth
everywhere, all respective multipole moments must be zero.
(This does not hold for the multipole moments pertaining to the Bessel functions
of the second kind since $Y_{j+1/2}(r)$ is singular at $r=0$.) 
% Note that this is valid
%even for more general boundary conditions at the substrate derived from
%Eq.~(\ref{eq:fs_lc}).

The excess force on the system ``colloid and nematic--air interface" follows from
Eq.~(\ref{eq:df2}) and can be expressed as
\bea
 \label{eq:stress_twisted}
 \Delta F = - \frac{K}{2}\int_{S_{\rm sub}} dA\; (v_{,z}^2-2q_0\,v_{,z}+w_{,z}^2) \;,
\eea
utilizing the boundary conditions %and symmetries of 
for the solutions $v$ and $w$.
For the multipoles appearing in the expansion of $v$ the method described in 
Subsec.~\ref{sec:aligned}
may be used. In order to obtain  the quadrupolar contributions to $w$ we perform the summation over the
image multipoles and the integration over the substrate surface numerically. Due to the
slow decay of $Y_{5/2}(r) \propto (\cos r)/r^{1/2}$ the stress integral is superficially
divergent. However, a detailed asymptotic analysis yields convergence with the result  
\bea
 -\Delta F = \pi \,K \left( \frac{3\zeta(3)}{2}\,\frac{P_v^2}{h^4} + \frac{5\zeta(5)}{12}\,
  \frac{Q_v^2}{h^6} + c_w\,(Q^Y_w)^2 \right) \;,\quad
\eea
where $c_w\approx 0.30$ has been determined numerically. 
%{\tt I have rewritten the following sentence:} 
Note that the linear contribution
%s in $P_v$ and $Q_v$ 
due to the term $\sim q_0\,v_{,z}$
in the stress tensor (Eq.~(\ref{eq:stress_twisted})) turns out to be zero, as can be 
easily checked by applying Gauss' theorem and the field equation $\Delta v = 0$.
Using the above considerations
concerning the magnitude of the multipole moments $P_v$, $Q_v$ and $Q^Y_w$ we obtain
\bea
 \label{eq:ftwist}
 -\frac{\Delta F}{K} \sim a_P \,\left(\frac{R}{h}\right)^{4+2\kappa} + 
  a_Q\, \left(\frac{R}{h}\right)^6\;,
\eea
where $a_P$ and $a_Q$ are numerical coefficients of order 1 and $\kappa$ has been
defined in Eq.~(\ref{eq:kappadef}). Again the excess force
on the colloid falls off rapidly with increasing film thickness such that the induced 
logarithmic capillary interaction (see Eq.~(\ref{eq:vmen_noniso})) remains very weak.
For the parameters characterizing the experimental system studied in Ref.~\cite{Lav04} 
($R/h\sim 10^{-1}$, $K/(\gamma R)\sim 10^{-4}$)
we find $V_{\rm men}(d) \sim$ \linebreak $10^{-11-4\kappa}\, k_B T\;\ln(R/d)$ 
which appears to be undetectably small.
Note that for $d<h$ the direct elastic repulsion remains essentially unchanged
because in this regime the leading term of the elastic interaction is given by the
repulsion between the two quadrupoles located at the colloid sites. In this case the image
quadrupoles can be neglected.

\section{Discussion and conclusion}
\label{sec:conclusion}

We have investigated the effective potential between two colloidal microspheres
of radius $R$
floating at asymptotically large distances $d$ on an interface between a nematic film 
of thickness $h$ and air (Fig.~\ref{fig:nem_exp}).
This effective potential is the sum of an elastic interaction caused by
the director distortions around the colloids and of a capillary interaction mediated
by surface deformations. We have analyzed the effective potential for large $d$ employing the
coarse--grained Frank free energy (within the one constant approximation)
for the director distortions and the
linearized Young--Laplace equation for the interface distortions.   

In the case of a macroscopically thick nematic film, the director deformation
around a single colloid is of quadrupolar type. Thus the induced elastic interaction
between two colloids at distance $d$ is repulsive and of quadrupolar type  $\propto d^{-5}$.
The capillary interaction is also repulsive and decays $\propto d^{-5}$ but it is much
weaker than the elastic interaction. This rapid decay is a consequence of the mechanical 
isolation of the system,
i.e., of the fact that the net force on the colloidal particles and
the surrounding interface vanishes.

A finite film thickness $h$ leads to a violation of the mechanical isolation of the
system ``colloid and interface". (The excess force is counteracted by the film substrate
such that the whole experimental system is of course in mechanical equilibrium.)
However, on the thermal energy scale the strength of the ensuing logarithmic 
capillary potential turns out to be very small. 
In the case of homeotropic boundary conditions on both sides of the nematic film
the strength is proportional to $(R/h)^{12}$ (see Eq.~(\ref{eq:fqq}) and the subsequent discussion).
 For  twisted boundary conditions
(parallel at the bottom substrate, perpendicular at the upper nematic--air interface)
there is a qualitatively different asymptotic behavior of the director field
due to a dipole contribution, which is induced by the broken rotational 
symmetry and vanishes for $h\to \infty$.
However, even for these boundary conditions 
%{\tt Here I have dropped the following sentence: ``since this dipole contribution induced by the finite film thickness
%must vanish for $h\to \infty$'', because the amplitude of the log would also vanish even if the dipole would not (Eq.~(43) with $\kappa=0$).}
the strength of the logarithmically varying 
capillary potential remains 
extremely
small, vanishing at least $ \propto (R/h)^8$ (see Eq.~(\ref{eq:ftwist}) and the subsequent discussion).
% and thus seemingly does not offer an explanation of 
%the recent experimental observations made in Ref.~\cite{Lav04}.

Thus our analysis based on the mechanical isolation of the
experimental system under consideration rules
out a significant attractive contribution $\sim \ln d$ of capillary type 
to the  effective potential
between two colloids at a nematic interface.
The amplitude of such a logarithmic contribution vanishes rapidly 
for large  film thickness $h$. Therefore, this effective pair potential 
does not provide a mechanism for the stability of isolated colloid clusters
at a nematic interface as reported in Ref.~\cite{Lav04}.
%We have demonstrated that the explanation offered in Ref.~\cite{Lav04}
%for the , based on an effective 
%pair potential between
%the colloids, is not valid.   Also the elastic contribution of dipolar type
%($\propto d^{-3}$) to the effective pair potential is not backed by our asymptotic 
%analysis.   
%Ref.~\cite{Lav04}
%derives an $h$--independent amplitude because only part of the forces
%contributing to the net mechanical balance are identified. 

Naturally one strives for other explanations for the observation of stable clusters
reported in Ref.~\cite{Lav04}. 
There the mutual center--to--center distance between 
neighboring colloids in the cluster has been found to be between $3R$ and $5R$, 
depending on the
radius $R$ of the colloids. If the cluster stability is attributed to a minimum
in the effective pair potential at this range $3R \dots 5R$ of distances, 
the applicability of the above asymptotic considerations at such distances is doubtful 
for both the elastic and capillary
contributions to the effective pair potential: 
\begin{itemize}
 \item {\em Elastic part:} For the single--colloid problem, the asymptotic, quadrupolar
  form of the director field (Eq.~(\ref{eq:quadrupole})) is based on the assumption
  that the deviations of the director from the preferred direction $\vect e_z$ are small: \linebreak
  $|n_{x[y]}| \ll 1$. At a radial distance $\rho$ from the center
  of the colloid this implies $Q/\rho^3 \ll 1$, which seems to be fulfilled for
  the dimensional estimate for the quadrupole moment $Q = O(R^3)$ and for the
  distances $\rho = 3R \dots 5R$ under discussion. However, the absolute magnitude of the
  director deformations is not fixed by these arguments, so that it might be
  worthwhile to determine the minimum of the Frank free energy 
  by a full numerical calculation in the presence of one or two colloids at
  the nematic--air interface, similar to Ref.~\cite{Sta99} where  the explicit 
  director solution around a single colloid in the bulk has been determined.   
  Our analysis indicates that 
  only if the multipolar expansion of the director field around a single trapped colloid
  fails for distances $\rho = 3R \dots 5R$, there is a chance for attractive elastic
  interactions between two colloids at these distances. 
  Failure of the multipole expansion at these intermediate 
  distances would point to a strong deviation of the magnitude of the 
  multipole moments $M_n$ of order $n$ from the dimensional estimate
  $M_n \sim R^{n+1}$. Thus it is certainly worthwhile to determine
  multipole renormalizations for bulk and interface configurations, taking into
  account the full nonlinearity of the director field equations.

 \item {\em Capillary part:} Whenever the colloid--induced nematic stress deviates from the
  asymptotic, quadrupolar form (Eq.~(\ref{eq:pizz_single})), corrections
  to the asymptotic capillary potential will
  arise. However, due to the smallness of the dimensionless
  force $\varepsilon_F$ (see Eq.~(\ref{eq:epsdef}), $\varepsilon_F \propto K/(\gamma R) \approx 10^{-3}$) 
  the perturbative treatment of the ensuing capillary deformations
  is justified. Therefore, one of our main findings, i.e., that the magnitude
  of the capillary potential ($\propto \varepsilon_F^2$) is always smaller
  than the magnitude of the elastic interaction ($\propto |\varepsilon_F|$), 
  is likely to hold also for elastic stresses deviating from the
  asymptotic limit.
%{\tt In case we decide to answer also the claim with the torques by
%  Lavrentovich et al.}: 

Another possible, subtle effect related to the 
anisotropy inherent to nematic phases
allows the substrate to exert also a torque on the colloidal particles
and the surrounding interface. This case can be studied similarly
to what we have done here. Actually, the result can be obtained 
straightforwardly if the analogy of capillary deformation with 2D
electrostatics is employed (see, e.g., Ref.~\cite{Dom06}): The net
force creates ``capillary monopoles'' $Q_{\rm cap}$ with an interaction
energy $V_{\rm men}(d) = Q_{\rm cap}^2/(2\pi\gamma)\, \ln( C\lambda/d)$ 
(see Eq.~(\ref{eq:vmen_noniso})). A net torque creates
``capillary dipoles'' ${\bf P}_{\rm cap}$ and the interaction energy
between them is $V_{\rm men}(d) \sim (|{\bf P}_{\rm cap}|/d)^2$. This decay is
sufficiently
slow to dominate asymptotically the
nematic--mediated repulsion. However, dimensional analysis yields $P_{\rm cap}
\sim R\, Q_{\rm cap}$ so that the amplitude of
this capillary energy, which must vanish for large film thicknesses $h\to\infty$, 
has also a
too small numerical value to explain the experimental results reported in Ref.~\cite{Lav04}.
\end{itemize}
%
%First, we want to emphasize that our strictly asymptotic
%considerations cannot exclude the possibility of stronger,
%non--logarithmical capillary attractions when the colloids are close
%to each other.
%
%Actually, we obtain indications that beyond the small--deformation
%regime of validity of our calculations, a capillary attraction might
%arise which would dominate asymptotically over the nematic--mediated
%repulsion. There is, however, no evidence that the experiments
%conducted in Ref.~\cite{Lav04} involve large interfacial deformations.

We recall that for simplicity our calculations have been carried out in the ``strong
anchoring'' limit. ``Weak anchoring'' would lead to a
different director field, and thus to different values of the
dimensionless forces $\varepsilon_F$ and $\varepsilon_\pi$ (see
Eq.~(\ref{eq:epsdef})). It would also lead to an explicit contribution to the
expression~(\ref{eq:F}) for the
free energy accounting for the anchoring ``wetting''
energies~(Eq.~(\ref{eq:fs_lc})), which in the ``strong
anchoring'' limit considered so far were subsumed in a (quantitatively
negligible) additive renormalization of the surface tensions. The consequences of 
this more general case of no strong anchoring have to
be explored yet, but we do not expect that this alters our
conclusion that a logarithmically varying capillary attraction is
ruled out by mechanical isolation ($\varepsilon_F = \varepsilon_\pi$),
because this latter result is based on very general principles.

Finally we note that for a two--dimensional system, the interactions between 
two upright circular colloids trapped at the nematic--isotropic interface have been
studied numerically \cite{Tas05} using a Landau--de Gennes free energy. 
In three dimensions, this corresponds to
long, cylindrical colloids which are aligned side--to--side at the interface.
For boundary conditions that yield a similar, quadrupolar behavior of the director 
field around a single colloid \cite{Pet98}, 
the effective interaction is found to be repulsive
and consistent with a power--law decay (for intermediate distance $d/R$ up to 7). 
In this particular two--dimensional situation, the repulsive interactions appear
to be longer--ranged $\propto d^{-1}$ and in the numerical results 
for the director field there is no trace of a sizeable interface
deformation which would lead to capillary attractions.

In summary, we have presented several arguments 
% that point to the
% absence of an effective potential between two colloids at a nematic interface
% which consists of a repulsive elastic and an attractive capillary part. 
which rule out an asymptotic attraction of capillary origin in the
effective interaction potential between two colloids at a nematic interface.
If one discards effective pair potentials as the source of the
stability of colloidal mesostructures as the ones found in Ref.~\cite{Lav04}, 
the question arises whether this stability is a consequence
of genuine many--body effects \cite{Lav07a}. Nematic surfaces (without colloids)
can stabilize regular patterns of surface defects \cite{deG74}. If these  defects
persist in the presence of colloids, the experimentally observed regular order of 
the colloids might
be attributed to them. This issue calls for further experimental and theoretical
investigations.

{\bf Acknowledgment:}
We acknowledge useful discussions with S.~Chernyshuk, O.~Lavrentovich,
B.~Lev, P.~Patr\'icio, and J.~M.~Romero--Enrique. M.~O. thanks the German 
Science Foundation (DFG) for financial
support through the Collaborative Research Centre (SFB-TR6) 
``Colloids in External Fields", project section D6-NWG.
A.~D. acknowledges financial support from the Junta de Andaluc{\'\i}a
(Spain) through the program ``Retorno de Investigadores''.

\begin{appendix}

\section{Corrections to the strong anchoring limit}

\label{sec:app1}

Here we consider the asymptotic director field around a single colloid at the nematic--air
interface in the case of finite anchoring strength. For large distances from the colloid,
the overall deviations from the preferred director $\vect n =(n_1,n_2,n_3) 
= (0,0,1)$ are small
and the total free energy of the nematic phase is given by the harmonic approximation
for the bulk part (Eq.~(\ref{eq:fasy})) and the Poulini expression for the surface part:
\bea
  {\cal F}_{\rm ne} & = & \frac{K}{2} \spaceint{V_{\rm ne}}{r} 
     \sum_{i=1,2} \left( \nabla n_i \right)^2 +
     \frac{W_1}{2} \int_{A_{\rm air-ne}}\!\!\! dA \;
   (\vect n \cdot \vect e_A)^2 \nonumber  \\
   & =& \frac{K}{2} \spaceint{V_{\rm ne}}{r} \left( \left( \nabla n_1 \right)^2 +
  \left( \nabla n_2 \right)^2 \right) + 
 \nonumber \\ &&  
  \frac{W_1}{2} \int_{A_{\rm air-ne}}\!\!\! dA \;
  (1-n_1^2-n_2^2)  \;.
\eea   
The second equation holds because the normal $\vect e_A$ of the nematic--air interface
$A_{\rm air-ne}$ is parallel to the $z$--direction and $\vect n^2=1$. 
For particles with radii $R$ in the $\mu$m range, such as those investigated 
in Ref.~\cite{Lav04}, 
the dimensionless parameter $\alpha = K/(|W_1| R)$ is smaller than 1. Minimizing
the free energy renders the Laplace equation in the bulk,
\bea
   \Delta n_i = 0 \qquad (i=1,2)\,,
\eea 
supplemented with the Robin boundary condition:
\bea
%     K \frac{\partial}{\partial z} n_i (x,y,z=0) - W_1 n_i (x,y,z=0) & = & 0 
%    \quad \to \nonumber \\
   \alpha R\, \frac{\partial}{\partial z} n_i (x,y,z=0) - n_i (x,y,z=0) & = & 0
  \quad (i=1,2) \;. \nonumber \\
 \label{eq:boun}
\eea
The general, asymptotic solution  is given by the multipole {\em ansatz} 
(Eq.~(\ref{eq:es_mul}))
\bea
  n_i &=& q_i\,\frac{1}{r} + \sum_{\beta=1}^3 P_{i\beta}\,\frac{r_\beta}{r^3} +  
      \sum_{\beta,\gamma=1}^3 Q_{i\beta\gamma}\,\frac{r_\beta\,r_\gamma}{r^5}
 + \dots
\eea
with the requirement of rotational covariance around the $z$--axis. In view of
the boundary condition (Eq.~(\ref{eq:boun})) we note that the derivative
in Eq.~(\ref{eq:boun}) applied to a multipole of order $m$ generates 
a multipole term of order
$m+1$. Thus, to leading order in $1/r$, the boundary condition can only be met if
the leading multipole ($n_i^{\ell}$) fulfills the strong anchoring condition
$n_i^{\ell}(z=0)=0$ and is accompanied by a subleading multipole ($n_i^{s\ell}$)
which is connected to the leading multipole through the condition
$\alpha R\, \partial_z n_i^{\ell}(z=0) = n_i^{s\ell} (z=0)$.
Both the monopole and the dipole do not fulfill the strong anchoring condition
$n_i^{\ell}(z=0)=0$.
Therefore, as before, we find that the multipole expansion starts with the quadrupole solution
and that it is necessarily accompanied by a hexapole:
\bea
 \label{eq:app_sol1}
  n_i & = & n_i^{\ell} + n_i^{s\ell} \;, \\
  n_i^{\ell} &=& Q\,\frac{z r_i}{r^5} + Q^{\rm mag} \epsilon_{i\alpha 3}\, \frac{z r_\alpha}{r^5}\;, \\
%  \mbox{and} \quad 
  n_i^{s\ell} &=& \alpha QR\, \frac{r_i(r^2-5z^2)}{r^7} +
               \alpha Q^{\rm mag}R \epsilon_{i\alpha 3}\, \frac{r_\alpha(r^2-5z^2)}{r^7}
  \;. \nonumber \\
 \label{eq:app_sol2}
\eea
Since $\partial_z n_i^{s\ell}(z=0)=0$, this solution 
(Eqs.~(\ref{eq:app_sol1})--(\ref{eq:app_sol2})) exactly satisfies the boundary condition in 
Eq.~(\ref{eq:boun}). Thus
the asymptotically dominant director field consists of a quadrupole--hexapole
superposition. 
Furthermore, the magnitude of the accompanying hexapole moment is small due to
the factor $\alpha$, therefore the results reported before for strong anchoring 
(based on the
leading quadrupole only) are unaffected by a finite anchoring strength at the nematic--air
interface.

\section{Force balance in a general configuration} 

\label{sec:app2}

\begin{figure}[t]
 \begin{center}
  \epsfig{file=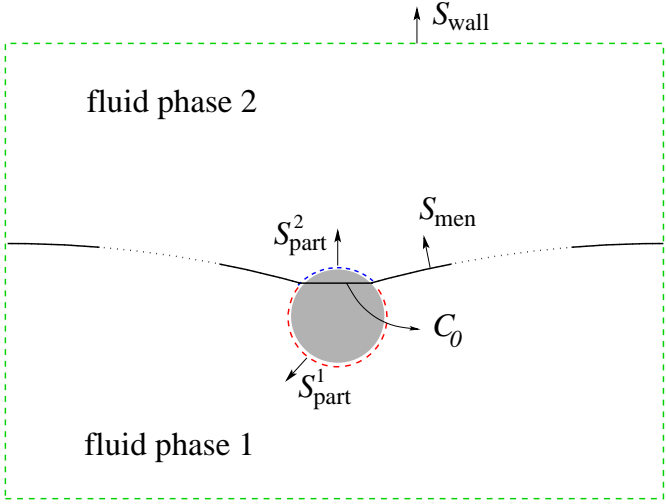, width=\columnwidth}
 \end{center}
 \caption{ 
   \label{fig:forcebalance}
   A general configuration of the particle and the fluid interface.
   The arrows indicate the orientation of the corresponding surfaces:
   $S_{\rm men}$ is the interface, $S^{1(2)}_{\rm part}$ is the
   surface of the particle in contact with the lower (upper) fluid
   phase, $S_{\rm wall}$ is the surface of the container of the system
   (sketched here as a quadrangular box for simplicity). The
   three--phase contact line between the particle and the interface is
   denoted as $C_0$.}
\end{figure}

For the benefit of the reader, we discuss in this Appendix some
previous results \cite{Dom05,Dom06} concerning the  force balance of a general
equilibrium configuration and demonstrate how the amplitude of an 
asymptotic, logarithmically varying interfacial deformation is
determined solely by this mechanical condition of force balance.

Figure~\ref{fig:forcebalance} represents a colloidal particle in equilibrium at
the deformed interface; in general the deformation is not small. 
The condition of mechanical equilibrium implies that locally
%not only that the net force on the whole system , but also 
the net force on any part of the system must vanish. Thus one
has:
\begin{enumerate}
\item Each of the fluid phases is in equilibrium. 
%The lower (nematic) phase 
%   , so that the integral of the
%   stress ${\mathbf \Pi}$ over the surface enclosing the phase must
%   vanish:
%   \begin{equation}
%     \int_{S_{\rm men} \cup S^{\rm wall}_{1} \cup (\mbox{}-S_1)}
%     d{\bf A} \cdot {\mathbf \Pi} = {\bf 0} .
%   \end{equation}
  We introduce the %(elastic and hydrostatic)
  forces exerted {\em by} each fluid phase  on the particle, on the whole
  fluid meniscus, and on the wall of the container as
  \bea
  \label{eq:fpart}
  {\mathbf F}_{\rm part}^{1(2)} & := & \int_{S_{\rm part}^{1(2)}} d{\bf A} \cdot {\mathbf \Pi}^{1(2)} , \\
  \label{eq:fint}
  {\mathbf F}_{\rm men}^{1(2)} & := & \mbox{} -(+) \int_{S_{\rm men}} d{\bf A} \cdot {\mathbf \Pi}^{1(2)} , \\
  {\mathbf F}_{\rm wall}^{1(2)} & := & \mbox{} - \int_{S_{\rm wall}^{1(2)}} d{\bf A} \cdot {\mathbf \Pi}^{1(2)} ,  
  \eea
  respectively, in terms of the stress tensor ${\bf \Pi}^{1(2)}({\bf r})$ in each
  fluid phase with due account for the orientation of the surfaces
  (see Fig.~\ref{fig:forcebalance}). The superscript $^{1(2)}$
  indicates the lower (upper) phase in Fig.~\ref{fig:forcebalance}. 
  The condition of mechanical equilibrium
  of each phase under the influence of these three forces reads
  \begin{equation}
    \label{eq:netFfluid}
    {\mathbf F}_{\rm part}^{1(2)} + {\mathbf F}_{\rm men}^{1(2)} +
    {\mathbf F}_{\rm wall}^{1(2)} = {\bf 0} .
  \end{equation}

  The total force exerted {\em by} the fluids on the particle is
  ${\mathbf F}_{\rm part} := {\mathbf F}_{\rm part}^{1} + {\mathbf
    F}_{\rm part}^{2}$, and on the meniscus it is ${\mathbf F}_{\rm
    men} := {\mathbf F}_{\rm men}^{1} + {\mathbf F}_{\rm men}^{2}$.
  The expressions for these forces reduce to those given in Eqs.~(\ref{eq:Fdef}) and
  (\ref{eq:Fpi}), respectively, upon evaluation in the reference 
   configuration depicted
  in Fig.~\ref{fig:ref}.
  %
%   {\tt Note: strictly speaking, Eq.(\ref{eq:Fpi}) should be defined
%     with the opposite sign}
  
  If the condition of mechanical equilibrium is applied 
  locally to an infinitesimal volume in
  the bulk of each of the fluid phases, it turns into $\nabla\cdot{\bf
    \Pi}^{1(2)} = {\bf 0}$. In the nematic phase this condition yields
  the field equations determining the director field.

\item The particle is in mechanical
  equilibrium under the combined action of ${\mathbf F}_{\rm part}$
  and the tension exerted {\em by} the interface on the particle at the
  three--phase contact line $C_{0}$. This tension can be expressed in terms of a line
  integral involving the surface tension $\gamma$:
  \begin{equation}
    {\mathbf F}_{\rm contact} := \mbox{} - \gamma \oint_{C_0} d\ell 
    \,{\mathbf e}_c , %\times {\mathbf e}_n ,
  \end{equation}
  where ${\mathbf e}_c$ is the unit vector tangent to the interface,
  normal to the contact line, and oriented towards the particle side. 
  Therefore, the condition of mechanical equilibrium reads
  \begin{equation}
    \label{eq:netFpart}
    {\mathbf F}_{\rm part} + {\mathbf F}_{\rm contact} = {\bf 0} .
  \end{equation}    

\item Any piece $S_{\rm int} \subset S_{\rm men}$ of the fluid
  interface is in mechanical equilibrium. The force on $S_{\rm int}$
  exerted by the fluid phases and the tension exerted on this piece at
  its boundary, $\partial S_{\rm int}$, are balanced:
  \begin{equation}
    \label{eq:netFint}
    \int_{S_{\rm int}} d{\bf A} \cdot 
    [{\mathbf \Pi}^2 - {\mathbf \Pi}^1]
    + \gamma \oint_{\partial S_{\rm int}} d\ell \,{\mathbf e}_c 
    = {\bf 0} ,
  \end{equation}
  with ${\mathbf e}_c$ oriented towards the exterior of $S_{\rm int}$.
  If this expression is applied locally to an infinitesimal piece of
  interface, it turns into an equation for the interfacial deformation
  relating the mean curvature of the interface to the pressure jump
  accross it. If, in addition, the interface deviates only slightly
  from a flat interface (identified with the plane $z=0$), this equation reduces
  in turn to the well--known equation~(\ref{eq:linearYLeq}) below
  %$\gamma \nabla^2 u = \Pi_{zz}^1 - \Pi_{zz}^2$ 
  for the local height $u(x,y)$ over the plane $z=0$.

\end{enumerate}

The net force balance of the whole system follows %of course 
from the three separate balance conditions~(\ref{eq:netFfluid}),
(\ref{eq:netFpart}), and (\ref{eq:netFint}): with $S_{\rm int} =
S_{\rm men}$ (so that $\partial S_{\rm int} = C_0 \cup C_{\rm wall}$;
$C_{\rm wall}$ is the three--phase contact line between phase 1, phase 2, and the
container enclosing the system),
one finds that, as expected, in equilibrium the net force of the outer environment on
the system must vanish\footnote{The reasoning can be
  easily generalized to the case that in addition to the
  short--ranged influence of the wall there are also external fields
  (gravity, electric force) acting on any part of the system.}:
% on the fluid phases by the wall of the container
% must be balanced by a tension at the contact line of the fluid
% interface with the wall
\begin{equation}
  \label{eq:netFtotal}
  {\mathbf F}_{\rm wall}^{1} + {\mathbf F}_{\rm wall}^{2} -
  \gamma \oint_{C_{\rm wall}} d\ell \,{\mathbf e}_c = {\bf 0} ,
\end{equation}
where ${\bf e}_c$ points to the exterior of the system.

The condition of mechanical equilibrium can be applied advantageously
to obtain a precise statement about the amplitude of an
interfacial deformation $u(\rho,\phi)$ varying logarithmically with
the lateral distance $\rho \gg R$ from the particle  with radius $R$. 
Far away from the particle, interface deformations and their gradients are
small, so that the linearized equation holds.
%Since a logarithmic dependence
%is a solution of the 2D Laplace equation, it can only arise provided
%there is a region where the deformation from a flat interface is small
%and accordingly the linearized equation holds.
% and there
% is no pressure force deforming the interface, so that $\gamma \nabla^2
% u =0$. Physically, this requires a clear spatial separation between
% the regions where the forces deforming the interface are applied, in
% our case the neighbourhood of the particle and that one of the wall.
Thus there is a distance $\xi$ beyond which the linear theory is applicable:
\begin{equation}
  \label{eq:linearYLeq}
  \gamma \nabla^2 u (\vect r) = \Pi_{zz}^1 (\vect r) - \Pi_{zz}^2(\vect r)  \qquad
  ({\bf r} \in S_{\rm ext}) \;,  %\xi < \rho ,
\end{equation}
where the piece of interface $S_{\rm ext}$ is enclosed by the
circle $\rho=\xi$ and the line $C_{\rm wall}$.  The general solution 
to this inhomogeneous Laplace equation can be written as
\begin{eqnarray}
  \label{eq:generalu}
  u(\rho,\phi) & = & A_0 + B_0 \ln\frac{\xi}{\rho} +  \\ && 
  \sum_{m=1}^{+\infty} \left[ A_m \left(\frac{\rho}{\xi}\right)^m 
    + B_m \left(\frac{\xi}{\rho}\right)^m \right] \cos m(\phi-\phi_m) +
   \nonumber\\
  & &  \frac{1}{2\pi\gamma} \int_{S_{\rm ext}} \!\!\! d\phi' d\rho' \rho'\;
  [\Pi_{zz}^1({\bf r}') - \Pi_{zz}^2({\bf r}')] \, 
  \ln\frac{|{\bf r}-{\bf r}'|}{\rho}\; . \nonumber
\end{eqnarray}
The fixed values of the constants $A_m$, $B_m$, and $\phi_m$ are determined by
the boundary conditions. This expression reduces to
Eq.~(\ref{eq:sol1}) in the particular case of rotational symmetry and
a wall located at infinity.

We can apply Eq.~(\ref{eq:netFint}) to the piece $S_{\rm int}=S_{\rm
  men}\backslash S_{\rm ext}$ enclosed by the contact line $C_0$
and the circle $\rho=\xi$. Invoking Eq.~(\ref{eq:netFpart}), one has
\bea
  {\bf F}_{\rm part} + \int_{S_{\rm int}} d{\bf A} \cdot [{\mathbf \Pi}^2 - {\mathbf \Pi}^1]
  &=& \mbox{} - \gamma \oint_{\rho = \xi} d\ell \,{\mathbf e}_c
  \\ \nonumber
  &\approx& \mbox{} - \gamma \, {\bf e}_z \int_0^{2\pi} d\phi \; \rho 
  \left.\frac{\partial u}{\partial \rho} \right|_{\rho=\xi} ,
\eea
where the last, approximate equality involves the leading order term in
$\nabla u$ of the line integral. (This approximation is justified because the interface 
deviates only slightly from a flat interface at the circle $\rho=\xi$.) 
Evaluating this latter term for the general
solution~(\ref{eq:generalu}) one finally finds that the amplitude of
the logarithmic term in Eq.~(\ref{eq:generalu}) is proportional to the
force exerted by the upper and the lower fluid on the particle and on the meniscus:
\begin{equation}
  \label{eq:amplitudelog}
  B_0 = \frac{1}{2\pi\gamma} {\bf e}_z \cdot \left[ 
    {\bf F}_{\rm part} + {\bf F}_{\rm men} \right] .
\end{equation}
In obtaining Eq.~(\ref{eq:amplitudelog}) we have used that
in the region $S_{\rm ext}$ (where the interface is almost flat)
one has to leading order
\begin{equation}
  \int_{S_{\rm ext}} d{\bf A} \cdot [{\mathbf \Pi}^2 - {\mathbf \Pi}^1]
  \cdot {\bf e}_z
  \approx 
  \int_{S_{\rm ext}} d\phi\, d\rho\, \rho \;
  [\Pi_{zz}^2({\bf r}) - \Pi_{zz}^1({\bf r})] .
\end{equation}
In conclusion, if the stress $\Pi_{zz}^1({\bf r}) - \Pi_{zz}^2({\bf
  r})$ decays sufficiently fast with the distance $\rho$ from the
particle, the solution in Eq.~(\ref{eq:generalu}) will be dominated
asymptotically by the logarithm with an amplitude given by
Eq.~(\ref{eq:amplitudelog}), provided this amplitude does not
vanish \footnote{To which extent this asymptotic regime is actually observable in
  a particular experimental realization depends on the precise
  functional form of ${\bf \Pi}^{1(2)}({\bf r})$ and the values of the
  constant parameters $A_m$, $B_m$, and $\phi_m$ entering the solution given by
  Eq.~(\ref{eq:generalu}).}. As one can infer from
Eq.~(\ref{eq:netFfluid}), this latter condition means physically that
the walls exert a non-vanishing force: a
logarithmic term can only arise if mechanical isolation of the system ``colloid + interface'' is violated.
(Note that Eq.~(\ref{eq:amplitudelog}) follows actually from
Eq.~(\ref{eq:netFtotal}) if the interfacial deformations are small.)

We emphasize the generality of this result: it only requires that the
interface departs slightly from a flat one for distances sufficiently far
from the particle. (Otherwise it is actually not useful to
speak of a logarithmically varying deformation to begin with.) In
particular, the interfacial deformation close to the particle may be
arbitrarily large, the particle itself need not be perfectly spherical,
or it may even consist of a many--body configuration lacking any kind
of symmetry.

\end{appendix}

\end{document}